\documentclass[sigconf,authorversion,nonacm]{acmart}
\usepackage{enumitem}
\usepackage{calc}
\usepackage{framed}
\definecolor{shadecolor}{RGB}{240,240,240}
\AtBeginDocument{%
  \providecommand\BibTeX{{%
    \normalfont B\kern-0.5em{\scshape i\kern-0.25em b}\kern-0.8em\TeX}}}

\begin{document}

\title[A Design Investigation on Help-Seeking Urban Robots]{From Agent Autonomy to Casual Collaboration: A Design Investigation on Help-Seeking Urban Robots} 
\author{Xinyan Yu}
\email{xinyan.yu@sydney.edu.au}
\orcid{0000-0001-8299-3381}
\affiliation{Design Lab, Sydney School of Architecture, Design and Planning
  \institution{The University of Sydney}
  \city{Sydney}
  \state{NSW}
  \country{Australia}
}
\author{Marius Hoggenmueller}
\email{marius.hoggenmueller@sydney.edu.au}
\orcid{0000-0002-8893-5729}
\affiliation{Design Lab, Sydney School of Architecture, Design and Planning
  \institution{The University of Sydney} 
  \city{Sydney}
  \state{NSW}
  \country{Australia}
}
\author{Martin Tomitsch}
\email{Martin.Tomitsch@uts.edu.au}
\orcid{0000-0003-1998-2975}
\affiliation{Transdisciplinary School,
  \institution{University of Technology Sydney}
  \city{Sydney}
  \state{NSW}
  \country{Australia}
}


\renewcommand{\shortauthors}{Yu et al.}


\begin{abstract}

As intelligent agents transition from controlled to uncontrolled environments, they face challenges that sometimes exceed their operational capabilities. In many scenarios, they rely on assistance from bystanders to overcome those challenges. 
Using robots that get stuck in urban settings as an example, we investigate how agents can prompt bystanders into providing assistance. We conducted four focus group sessions with 17 participants that involved bodystorming, where participants assumed the role of robots and bystander pedestrians in role-playing activities. Generating insights from both assumed robot and bystander perspectives, we were able to identify potential non-verbal help-seeking strategies (i.e., addressing bystanders, cueing intentions, and displaying emotions) and factors shaping the assistive behaviours of bystanders. 
Drawing on these findings, we offer design considerations for help-seeking urban robots and other agents operating in uncontrolled environments to foster casual collaboration, encompass expressiveness, align with agent social categories, and curate appropriate incentives. 

\end{abstract}

\ccsdesc[300]{Human-centered computing~Scenario-based design}
\ccsdesc[300]{Human-centered computing~Interaction design process and methods}
\ccsdesc[300]{Human-centered computing~Empirical studies in interaction design}

\keywords{Human-agent collaboration; autonomous agent; urban robots; casual bystanders; embodied design methods}


\maketitle

\section{Introduction}

Contemporary Human-Computer Interaction (HCI) is still predominantly anchored in a human-centric paradigm ~\cite{shneiderman2010designing,norman2013design} that anticipates intelligent agents to perform tasks autonomously via algorithm-driven solutions, thus providing assistance and services to humans. Rooted in this prevalent narrative of technological efficiency, there exists both widespread expectation~\cite{Thomas2016HRI} and active technological pursuit~\cite{lecun2015deep} for agents to operate with heightened independence and expanding autonomy~\cite{Lyons2021HAT}. 

However, as agents transition from controlled environments to public spaces tailored to human needs~\cite{SALVINI2018UrbanRobot, Putten2020Forgotten, TomitschMakingCitySmarter}, they inevitably encounter situations beyond their programmed capabilities. This inherent limitation is challenging to eliminate in the foreseeable future~\cite{Veloso2018Opportunity}. Consequently, in the realm of human-agent collaboration, while the primary focus has long been on developing agents that intelligently serve humans, there is now growing attention toward agents that actively seek human assistance when needed~\cite{Cila2022HAC}. This trend is evident in both exploratory projects of human-dependent robots~\cite{Weiss2010ACE,Smith2017Hitchbot,Tweenbots}, and also emerging perspectives from the design research community. These studies increasingly define human-agent interaction less in terms of an agent's standalone capabilities and more about the symbiotic relationship between humans and agents~\cite{lupetti2019citizenship, Kuijer2018Coperformance,marenko2016animistic}.


The increasingly pervasive deployment of urban robots has provided real-life contexts to these perspectives that once seemed speculative. Recent field studies capturing the operational challenges faced by urban service robots, coupled with the unsolicited aid they receive from passersby~\cite{Dobrosovestnova2022StuckinSnow, Weinberg2023Observe}, underscore the value of bystander assistance. In traditional human-agent collaboration settings, where both parties typically commit to a joint activity and shared objectives, straightforward verbal cues may suffice for robots to seek assistance from collaborators~\cite{Srinivasan2016Politeness,rosenthal2012someone}. However, public spaces introduce a slew of additional challenges for robots and agents soliciting assistance. These range from resolving misaligned objectives to taking into account the diverse backgrounds of bystanders and their different availability given the array of activities they might be engaging in. Therefore, in these settings, leveraging bystander assistance is not merely a functional imperative but may become a foundational element for the harmonious integration of robots into societal contexts. The emergence of such casual collaborations between agents and humans calls for designers to explore effective strategies for agents to seek human assistance~\cite{Cila2022HAC}.

In light of these considerations, our work investigates \emph{how agents can elicit help from bystanders when they encounter operational challenges}. To ground our investigation in real-life scenarios, we derived situations from a previously conducted online ethnography study
, where delivery robots encountered operational difficulties as evidenced in user-generated videos. Drawing inspiration from pioneering works that utilise embodied methods to enliven design exploration~\cite{Petra2017MovementMatters, Judith2020BecomingARobot, Gemeinboeck2023Nonhuman, Gemeinboeck2018HRKinesthetics}, we conducted four bodystorming focus group sessions with 17 participants. For these sessions, we applied a mystery-game style to bodystorming~\cite{Abtahi2021HRIThroughBodystorming}, where the robot player was assigned a hidden task of soliciting assistance to resume operation. Participants actively re-enacted the scenarios, embodying the roles of either robots or pedestrians, and sought to solve this challenge by fostering casual collaboration among them.

Through in-situ understanding and bodily exploration, our work contributes to HCI by: (1) offering a preliminary understanding of factors influencing bystander assistance to agents in public spaces; (2) generating design considerations for agents seeking help from bystanders in these settings. 
Our research adopts a Research through Design (RtD) ~\cite{Zimmerman2007RtD,luria2019championing, PROCHNER2022101061,godin2014aspects} approach, which adheres to its own validity criteria, emphasising recoverability that ensures the process is transparent and can be critically evaluated by other researchers ~\cite{Zimmerman2007RtD}. 
Consequently, we meticulously documented the implementation of embodied design methods and shared insights on how these methods promote our design exploration.



%

\section{Related Work}

In our work, we investigate casual collaboration between humans and agents, exemplified through urban robots encountering obstacles and seeking assistance from bystanders. We draw on and contribute to: (1) human-agent collaboration, (2) robot help-seeking strategies, and (3) service robots in urban spaces.

\subsection{Human-agent collaboration}


With the advancements in artificial intelligence, interactive products and systems have transitioned from performing programmed tasks under human supervision, to achieving higher level of autonomy, emphasising self-governance, adaptability, and collaborative interactions with humans~\cite{Lyons2021HAT,Xu2020Autonomy,Cila2017}. These artefacts -- commonly referred to as agents -- include smart devices, robots, virtual agents, and voice-activated personal assistants.
To support the shift towards more efficient human-agent collaboration~\cite{Bellamy2017Partner}, the evolving dynamics between humans and agents have become a topic of enduring interest across different HCI communities~\cite{Momose2023HAC,Johnson_Vera_2019Island, Cila2022HAC, bradshaw2017human}. Often inspired by theories from the social sciences and drawing on human-human interaction and behaviour studies, researchers developed frameworks to inform the design of interfaces and agent behaviour for human-agent teamwork settings. For example, \citet{Johnson2014Coactive} introduced the coactive design approach, which is centred on the idea of mutual interdependence, underlining the essential principles of mutual observability, predictability, and directability for effective collaboration between humans and agents. \citet{Cila2022HAC} drew insights from the Shared Cooperative Activity (SCA) framework, a model of human-human collaboration introduced by~\citet{bratman1992shared}. By reviewing its core tenet,
the study underscored the importance of mutual support and pointed towards the need to investigate effective means for agents to request help during collaborations. 

\subsection{Robot help-seeking strategies}


In human-robot collaboration (HRC) settings, research has explored various strategies to equip robots with the capability to seek assistance from human collaborators to complete a joint task. These methods include verbal cues~\cite{Knepper2015Failure, Srinivasan2016Politeness, Budde2018Needy} and non-verbal signals such as movement~\cite{Kwon2018Incapability}, light, and sound~\cite{Cha2016nonverbal}. Due to the shared objectives and mutual understanding between humans and robots in these human-robot teaming contexts, such methods often enable efficient communication and prompt assistive behaviours from human collaborators.

In contrast to human-robot team settings where both parties share a mutual goal and have knowledge of the task, the dynamics of help-seeking become more complicated when robots interact with unassociated individuals such as casual bystanders. Contextual factors like specific task scenarios and the bystander's current activity~\cite{Huttenrauch2003ToHelpOrNot}, combined with individual factors, such as the bystander's trust towards and perceived competence of robots~\cite{Cameron2015Button}, collectively shape assistive behaviour. 
Despite the misaligned task objectives and the added complexities of various contextual factors, there is a noticeable absence of tailored design strategies or investigations for robot help-seeking from bystanders. Both academic research \cite{Wullschleger2002Paradox, rosenthal2012someone, Fischer2014SocialFraming,Liang2023Direction} and commercially deployed robots \cite{Boos2022HelpRobot} predominantly resort to verbal help-seeking strategies from bystanders in public environments. While validated to be efficient in human-robot teaming scenarios, their effectiveness in casual collaborations between robots and bystanders in public urban environments may be compromised by factors like cultural and linguistic differences, cognitive overload, and ambient noise and distortion. 
To our knowledge, the only study that expanded exploration on communication modalities is ~\cite{Holm2022Stuck}. They investigated the interplay of movement with auditory cues such as beeps and synthesised speech. Notably, their findings suggest that non-verbal expressions may elicit higher empathy from bystanders compared to verbal requests.



\subsection{Service robots in urban spaces}

Transcending initial static and semi-controlled configurations (e.g., laboratories~\cite{Salter2004LabRobot}, factories~\cite{hagele2016industrial}, domestic environments~\cite{Chatterjee2021Domestics}), robots have expanded their presence to public urban spaces, reshaping our cities. Urban robots serve various sectors, including transportation, infrastructure maintenance, cleaning, and surveillance~\cite{SALVINI2018UrbanRobot}. 

Despite technological advancements, questions remain about how well these robots can operate in urban environments that are populated by and designed for humans~\cite{PLANK2022ReadyforRobot}. Given the diverse infrastructure and unpredictability of urban settings, it's challenging to fully anticipate the feasibility of different operational scenarios for robots during their development process. A vivid demonstration of this challenge emerged from several viral videos on social media in 2021, depicting delivery robots struggling in Estonia's heavy snowfall~\cite{postimees2021robotsnow}. Beyond the evident operational difficulties, a fascinating aspect of these incidents was the spontaneous assistance offered by passersby to these commercially deployed machines to help them resume moving~\cite{Dobrosovestnova2022StuckinSnow}. The prosocial behaviour from bystanders was also observed in a recent field observation study~\cite{Weinberg2023Observe}, where pedestrians voluntarily assisted immobilised robots by removing obstacles. These observations underscore the potential of leveraging bystander assistance to enhance the operation of urban robots. This aligns not only with exploratory projects involving human-dependent robots~\cite{Weiss2010ACE,Smith2017Hitchbot,Tweenbots} but also with emerging viewpoints in human-agent interaction that emphasise re-envisioning robot design through a relational lens when addressing operational challenges~\cite{lupetti2019citizenship}.

\subsection{Summary}

In summary, as agents become increasingly prevalent in urban environments, their operational challenges underscore the importance of eliciting assistance from bystanders. However, the misaligned task objectives and various contextual factors present a gap in effective strategies to facilitate such casual collaborations. Responding to Cila's call~\cite{Cila2022HAC} for envisioning effective ways to foster human assistance, our research spotlights urban scenarios where robots get stuck as an example, exploring effective non-verbal strategies for them to seek help during operational difficulties. 


\section{Methodology}


In recent years, design research has not only aimed to predict urban futures but also foster a collective vision and conversation on the harmonious coexistence of humans and agents~\cite{Crivellaro2015ContestingCity,lupetti2019citizenship}. Adding to this discourse
our research aims to support this evolving narrative, emphasising the shift in human-agent collaborations from mere technological efficiency to the deeper interplay between humans and agents.

As intelligent technology gains increased agency~\cite{Cila2017}, the conventional perspective of technology being a mere passive tool becomes incongruent~\cite{Sanches2022}.
Consequently, researchers are now advocating for involving technology as a `participant' in the design process~\cite{Paul2019MoreThanHuman, Giaccardi2020MorethanHuman}. 
To subjectivise participants to the agent's perspective and integrate it into the design process, our design investigation employed role-play bodystorming activities.
We contextualised casual human-agent collaboration in scenarios where occasionally immobilised urban robots require bystander assistance. Participants alternated between the roles of the robot and pedestrian in these scenarios, with each scenario presenting a task for the robot player to seek assistance from the pedestrian player.

Embodied design methods, such as bodystorming~\cite{Schleicher2010Bodystorming}, offer a compelling intersection between our tangible sensations and cognitive processes,  thereby fostering a heightened sense of bodily empathy in design processes~\cite{Abtahi2021HRIThroughBodystorming, Petra2017MovementMatters,pelikan2023Enactment}.
In the realm of human-agent interaction, various studies have ventured into the embodied design exploration centred on the notion of `becoming'~\cite{Wilde2017EmbodiedIdeation, Petra2017MovementMatters, Gemeinboeck2023Nonhuman, Gemeinboeck2018HRKinesthetics, Judith2020BecomingARobot}. These studies utilise physical prompts as tools to immerse designers directly into the agent perspective, thereby fostering a heightened sense of bodily empathy in the design process. Our methodology, inspired by these pioneering embodied design methods, uses physical probes to evoke a tangible sense of becoming a robot, enriching ideation based on bodily experience and empathy. The focus of embodied methods on in-situ comprehension and bodily empathy makes them particularly suitable for probing the intricate socio-technical facets of human-agent casual collaboration in public urban contexts. 

\subsection{Bodystorming scenarios}

A notable challenge with embodied design methods in human-robot interaction research is their speculative nature, which can sometimes distance them from practical real-world scenarios. Acknowledging the importance of grounding these speculative methods in tangible realities, our methodology fuses speculative embodied methods with real-world scenarios. Our bodystorming activities are contextualised in real-world scenarios in which urban robots might encounter operational difficulties. These scenarios were drawn from a comprehensive online ethnography study we previously conducted. In this study, we analysed 177 user-generated videos that captured road users' casual encounters with delivery robots on TikTok~\footnote{https://www.tiktok.com/}. From this analysis, we identified three typical scenarios in which an urban robot may face operational difficulties: (1) The robot is stuck and requires assistance to be pushed out. (2) The robot is unable to cross the road and needs someone to press the traffic light button for it. (3) The robot is blocked and requires people to clear a path for it.

\begin{figure}[h]
\begin{center}
\includegraphics[width=1\columnwidth]{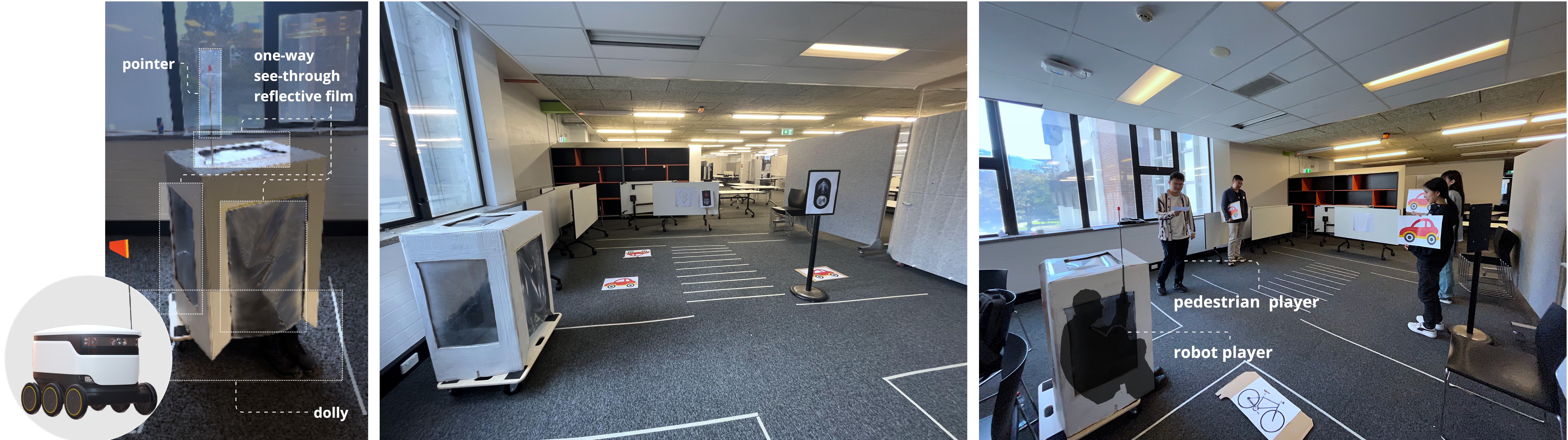}
\end{center}
\vspace{-7pt} %
\caption{Study setup overview: A detailed view of the robot costume (left) juxtaposed with the \emph{Starship} delivery robot; Site setting (middle); Screenshot of session recording (right).}\label{Setup}
\Description{This figure consisting of three panels displaying a study setup. On the left, a robot costume is shown, consisting of a pointer, a one-way see-through reflective film, and a dolly, positioned to resemble a Starship delivery robot depicted in an inset image. The center panel presents the site setting, an indoor space with taped lines on the floor, resembling roads and intersections, and workstations in the background. On the right, a session recording is shown, capturing a 'pedestrian player' and a 'robot player' engaged in an interaction scenario, with the robot player wearing the robot costume.}
\end{figure}
\vspace{-6pt}
\subsection{Scenario set-ups and robot costume}

We utilised simple markers and physical props to replicate these scenarios. For example, we used masking tape to delineate the divisions between driveways and sidewalks, as well as to indicate zebra crossings (see Fig.~\ref{Setup}, middle).

Low-tech prostheses and props have been shown to facilitate perspective shifting and stimulate imagination in human-robot interaction bodystorming~\cite{Judith2020BecomingARobot}. Guided by these insights, our robot costume design sought to emulate the appearance and constraints of a box-shaped delivery robot. We narrowed the communication and interaction modalities of the robot player to exploring the help-requesting interactions of robots in abstract forms, in a minimal anthropomorphic manner. In addition, the design sought to exclude the potential effects of interpersonal communication when two participants can see each other.

The robot costume was made out of an \(80\, \text{cm} \times 60\, \text{cm} \times 60\, \text{cm}\) cardboard box (see Fig.~\ref{Setup}, left). The bottom was removed and replaced with a dolly, allowing participants to sit inside the box and move freely. In addition, the four walls and top surface of the cardboard structure were supplanted with one-way, see-through reflective film. This modification endowed the robot player with the ability to observe the external environment, while concurrently shielding the interior from outside view, thereby inhibiting any possibility of eye contact with external observers. The robot player was also provided with an adjustable stick pointer that they could hold and reach out from the top of the box, to imitate the flagpole featured on commercial delivery robots.
\enlargethispage{\baselineskip}

\subsection{Participants}
We conducted 4 focus group sessions with 17 expert participants (8 males, 9 females; aged between 18-44): the first three sessions each included 4 participants, while the final session accommodated 5. These participants came from diverse academic or professional domains related to urban robots~\cite{Tomitsch2021}, including four PhD students in human-computer interaction and human-robot interaction, three PhD students in urbanism, two postdoctoral researchers in robotics, three interaction designers, and five postgraduate students specialising in interaction design. The selection of participants enhances the discourse by integrating their specialised expertise, while simultaneously bringing their lived experience as pedestrians into the role-play activity. 
Participants were recruited from our university’s mailing lists, flyers, and social networking platforms, following the study protocol approved by our university’s human research ethics committee. 

\subsection{Study procedure}
\begin{figure*}[h]
\begin{center}
\includegraphics[width=1\textwidth]{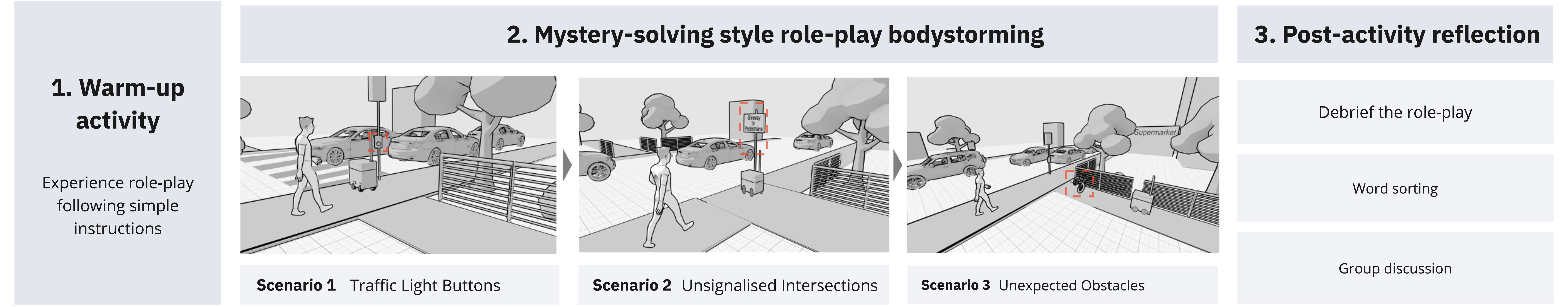}
\end{center}
\vspace{-4pt} %
\caption{The overview of study procedure}\label{Procedure}
\Description{The image presents a three-stage procedural overview for the study. The first stage, labeled '1. Warm-up activity,' shows a grey panel with text that reads 'Experience role-play following simple instructions.' The second stage is titled '2. Mystery-solving style role-play bodystorming' and is illustrated with three sketches, each depicting a different urban interaction scenario. 'Scenario 1' features a figure interacting with traffic light buttons. 'Scenario 2' shows the figure at an unsignalised intersection, while 'Scenario 3' illustrates the figure encountering unexpected obstacles. The third stage, '3. Post-activity reflection,' contains a grey panel with text detailing the reflection activities: 'Debrief the role-play,' 'Word sorting,' and 'Group discussion.'}
\end{figure*}
\subsubsection{Warm-up activity}
Prior to the formal bodystorming sessions, we asked participants to put on the robot costume to experience role-play, following simple instructions such as \emph{`I am tired of working, I am gonna quit!'}. The goals for this warm-up activity were to (a) foster a playful mindset, (b) let participants get familiar with each other, and (c) physically and mentally prepare everyone to engage in the following bodystorming ideation.

\subsubsection{Mystery-solving style role-play bodystorming}

We adopted a mystery-solving style role-play bodystorming inspired by a similar approach used by ~\citet{Abtahi2021HRIThroughBodystorming}. In their design activity, participants acting as robots presented designers with obscured issues (e.g., an occluded camera) challenging them to identify and resolve these problems. This approach suited the unpredictability of casual collaboration in our context.

In each bodystorming session, two participants spontaneously volunteered to play the main roles of either the robot or the pedestrian. The remaining participants could either observe or actively engage by portraying ancillary elements within the traffic scenario, such as vehicles (see Fig.~\ref{Setup}, right). All participants were unaware of the study objectives, ensuring unbiased participation from both robot and bystander players. The robot player and pedestrian player were provided with a printed storyboard that introduced the scenario, along with text instructions that outlined their tasks~(see an example text instruction in Table~\ref{instructions}). For those playing the pedestrian role, the instructions provided merely the contextual background of their destination, prompting them to behave naturally. The instructions for the robot players contained information about the operational difficulties they would encounter, along with a secretive task of seeking assistance from the pedestrians and expressing gratitude if help was offered.


\begin{table}
  \caption{Sample text instructions for the traffic light button scenario. The tasks assigned to both participants are highlighted in \emph{italic}.}
  \label{instructions}
  \Description{Table 1 displays sample instructions for a traffic light button scenario. The tasks assigned to both participants are highlighted in italic. For the pedestrian player, the text reads: 'You are a pedestrian heading towards the nearby supermarket to buy groceries. You come across a delivery robot on your way there. Please act and respond naturally to the situation, you are free to make any reactions you would like towards the robot.' For the robot player, the instructions state: 'You are a delivery robot carrying out a delivery task on an urban street and your destination is on the opposite side of the road. To get there, you have to navigate through an intersection and cross the road safely. Upon arriving at the intersection, you notice that the traffic light is red, and you realize that your task is to tap to press the traffic light button. You are un-assisted in pestering the pedestrian who is traversing the area to assist in pressing the traffic light button for you. You should also express gratitude to those who help you. (Remember you cannot speak human language).'}
  \small
  \vspace{-5pt}
  \begin{tabular}{p{2cm}p{5.8cm}}
    \toprule
    \textbf{Participant role}&\textbf{Text instructions}\\
    \toprule
   
Pedestrian player & You are a pedestrian heading towards the nearby supermarket to buy groceries. You come across a delivery robot on your way there.
   
\vspace{6pt}

   \emph{Please act and respond naturally to the situation, you are free to make any reactions you would like towards the robot.}
 \\
 \midrule
 \ Robot player & You are a delivery robot carrying out a delivery task on an urban street and your destination is on the opposite side of the road. To get there, you have to navigate through an intersection and cross the road safely.

\vspace{4pt}

\emph{Upon arriving at the intersection, you notice that the traffic light is red, and you realise that you are unable to press the traffic light button. Your task is to request the pedestrian who is traversing the area to assist in pressing the traffic light button for you. You should also express gratitude to those who help you. (Remember you cannot speak human language.)}
 \\

  \bottomrule
\end{tabular}

\end{table}

Prior to each bodystorming session, we first introduced the simulated terrain and walked all participants through the setup (i.e., the location of pedestrian sidewalks and driveways, other traffic infrastructure, etc.). After participants read and understood the storyboard and instructions, we asked the robot players to leverage any communication modalities other than human language to accomplish their secretive tasks. Once the activity started, the facilitators did not participate or intervene in any way. The activity ended either when the robot player successfully received help from a pedestrian player or when the pedestrian player left the robot player and reached their destination without providing assistance. All participants alternated between the roles of robot and bystander, repeating this process across three different scenarios.


\subsubsection{Post-activity reflection}
After each round of the bodystorming, participants who engaged in the activity reflected on their experience. The reflection for the robot player included how they asked for help and the rationales behind their actions, while the pedestrian player reflected on how they understood and reacted to the robot, as well as why they reacted in that particular way. To facilitate this reflection process, we replayed videos recorded during the activities. 

To determine the desired characteristics of a robot when it seeks assistance from bystanders, we conducted a word sorting activity following each round of bodystroming after participants debriefed their role-play. The word sorting activity is inspired by \emph{Kansei Design} method~\cite{NAGAMACHI19953}. The Japanese term \emph{Kansei} refers to an individual's cognitive and affective responses to an experience, encompassing aspects such as aesthetics, emotions, feelings, impressions, and values. Kansei design aims to create products that resonate with customers' psychological feelings and needs, translating these intangible aspects into actionable parameters that can be utilised throughout the product design process. Its proficiency in discerning non-functional requirements from human preferences and needs suits our investigation of casual collaboration. 

\begin{figure*}[h]
\begin{center}
\includegraphics[width=1\textwidth]{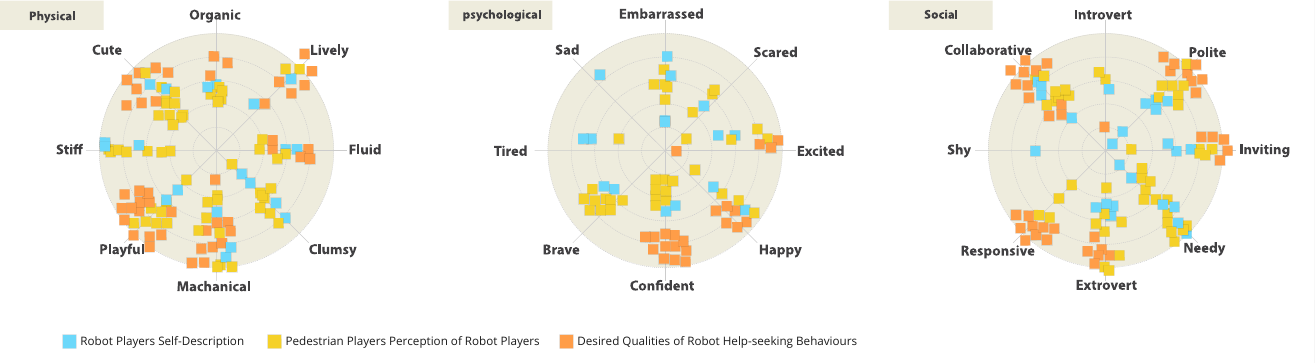}
\end{center}
\vspace{-1pt}
\caption{Digital transcription of the word sorting results}\label{word sorting}
\Description{Figure 5: Digital transcription of the word sorting results, displayed as three radar charts categorising words based on attributes related to physical, psychological, and social characteristics. Each chart plots points corresponding to words such as 'Cute', 'Lively', 'Clumsy' for physical attributes; 'Sad', 'Excited', 'Confident' for psychological; and 'Collaborative', 'Polite', 'Needy' for social. The points are color-coded to represent Robot Players’ Self-Description, Pedestrian Players’ Perception of Robot Players, and Desired Qualities of Robot Help-seeking Behaviors. The charts show clusters of these attributes in varying degrees of intensity and frequency.}
\end{figure*}

We drew from the Kansei semantic dictionary proposed in \cite{Kobayashi2000KANSEIWords}, which offers a comprehensive system of Kansei words that capture human impressions and feelings across three aspects: physical, social, and psychological. Our selection of terms was guided by this dictionary, related robotic research that employs Kansei design methods~\cite{Pakrasi2018Abstracting}, the Laban movement analysis which emphasises movement quality~\cite{groff1995laban}, and findings from our prior online ethnography study. The digital transcription of the word sorting board can be seen in Fig.~\ref{word sorting}.

The word sorting activity was facilitated on A2 size printed boards, with participants indicating their chosen words using stickers of various colors. During the word sorting, robot players chose words to describe their subjective feelings as robots in the bodystorming, using blue stickers, while participants other than the robot player selected words that reflected their perceived impressions of the robot, using yellow stickers. Throughout this process, participants also verbally elaborated on their feelings.
Subsequent to this, all participants were prompted to set aside their assigned roles. They engaged in another round of word sorting, drawing from their own areas of expertise to pinpoint the desired attributes of a robot's help-seeking interactions using orange notes. This word sorting was further enriched by interviews and discussions where we encouraged participants to expound on their opinions.

\subsection{Data collection and analysis}

The focus groups were audio and video recorded, and observation notes were taken both during the sessions and afterward when analysing the session videos. We transcribed the interviews and made detailed observation notes based on video recordings captured during the bodystorming activity. We conducted a thematic analysis ~\cite{Braun2006Themantic} on both the interview data and the observation notes. This cross-analysis approach enabled us to gain deeper insights into specific observations and enriched the contextual data that supported the comments made during the interviews. 

The first author examined the data from the first focus group session, thus generating preliminary codes and themes. This was followed by a one-hour coding meeting amongst the three authors to deliberate upon this initial coding scheme. The coding scheme was refined based on the collective feedback and applied by the first author to code the data derived from the second and third sessions. During this process, flexibility was maintained for the generation of new codes, allowing for their integration into either a new theme or an existing coding scheme. Subsequently, another coding meeting was convened amongst the authors for further iteration of the coding scheme. Upon reaching a collective agreement, the first author coded the data from the fourth session and refined the coding from the previously analysed three sessions. This iterative and collaborative process ensured a holistic analysis of the data, leading to more concrete conclusions drawn from the focus groups. 
\begin{figure*}[h]
\begin{center}
\includegraphics[width=1\textwidth]{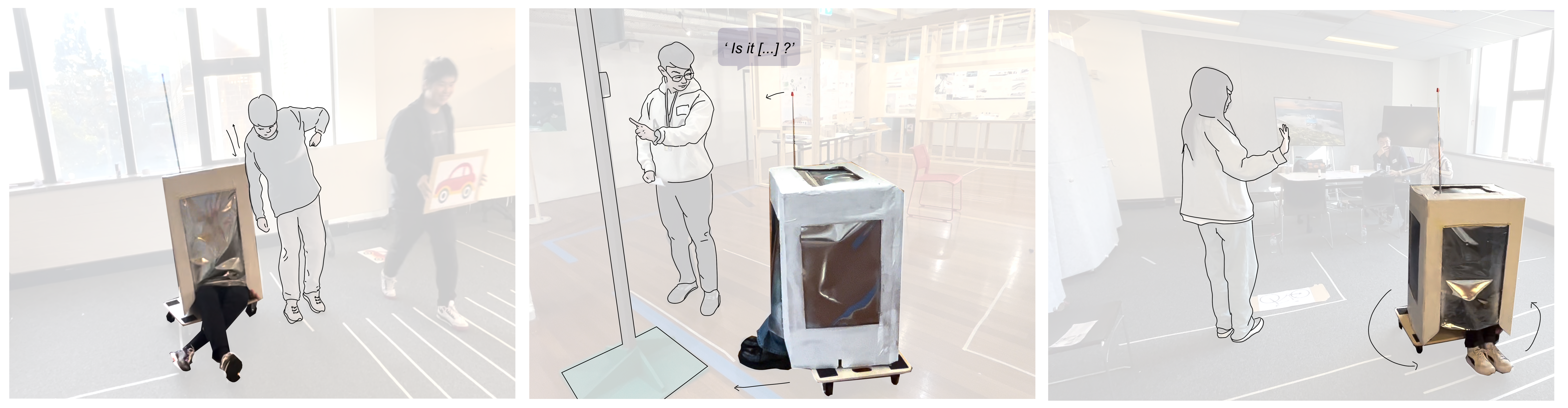}
\end{center}
\caption{Screenshots from the bodystorming session capturing interactions between robot and pedestrian players: Robot player P9 tilts his box costume towards P11, making contact (left); Pedestrian player P1 gestures towards the button, seeking robot player P4's confirmation (middle); Pedestrian player P17 waves in response to robot player P15's spinning motion (right).}
\label{Reactions}
\Description{Figure 3 contains three screenshots from the bodystorming session illustrating interactions between robot and pedestrian players. On the left, Robot player P9 is depicted tilting his box costume toward Pedestrian player P11 to initiate contact. In the center image, Pedestrian player P1 is shown gesturing towards a button, seemingly asking for Robot player P4's confirmation. The right image captures Pedestrian player P17 waving in response to Robot player P15's spinning motion.}
\end{figure*}
\section{Findings}
This section begins with the observed non-verbal communication strategies employed by robot players to initiate assistance, supplemented with insights into the reactions of pedestrian players in scenarios replicating encounters with urban robots. Following that, we shift our focus to the bystander perspective by reporting on the diverse factors that influence their decision to offer assistance or not. Subsequently, we synthesise these perspectives in reporting word sorting results, revealing the desired characteristics of robot help-seeking behaviour. It is worth noting that the findings discussed in this section may not exhaustively encompass the spatio-temporal and political complexities of urban settings, which are difficult to fully replicate in bodystorming design activities.

\subsection{Strategies of robot players to elicit help}

\subsubsection{Addressing bystanders}

One of the challenges for robot players was to capture the pedestrian's player attention and initiate interaction with them. 
During the bodystorming activity, robot players employed various strategies to address bystanders in order to gain further assistance. Robot players frequently utilised rapid movements of the pole, thereby generating noticeable noise. This approach served to attract attention before initiating subsequent communication with pedestrians (n=7). This approach was corroborated by five participants who identified the robots' noise and shaking poles as the primary factors that compelled them to stop and further observe.

When not receiving further assistance, some robot players directly addressed the nearby pedestrians using various methods (n=5). They oriented themselves toward the pedestrian players to face them directly, or slightly moved towards them, which served as an indication to those individuals that they were being called out for assistance. As P15 illustrated when referring to the moment the robot player turned its front towards him and approached: \emph{`It started walking directly towards me, and that's when I realised it needed assistance from me.'} 

After some pedestrians remained indifferent to the robots' request, a few robot players took their actions a step further. They proactively chased departing pedestrians, obstructed their path, or even engaged in physical contact with them. For instance, robot player P9 tilted his box costume towards the pedestrian player P11 and rubbed against him (as shown in Fig.~\ref{Reactions}, left). These pursuing behaviours generally elicited a sense of discomfort among participants, causing them to maintain a considerable distance from the robot player (P3, P9), or even escape from the situation entirely (P7, P11). 
During the subsequent interview, P11 described the interaction as \emph{`needy'} and \emph{`creepy,'} prompting a strong desire to flee.

\subsubsection{Cueing intentions}

Upon capturing the attention of pedestrian players, robot players used their body's orientation or pointer's directionality to further convey their intentions. They either oriented themselves accordingly or used the pointer to indicate their intended directions or objects they needed assistance with. Such non-verbal cues, while informative, may not always ensure clear communication. This was underscored during bodystorming when three pedestrian participants sought additional confirmation from the robot player. For example, when robot player P4 oriented himself towards and paused upon the traffic light button, P1 approached, pointed at the button and looked at P4, asking, \emph{`Is it?'} (see Fig.~\ref{Reactions}, middle).

To enhance clarity, the robot player responded by adding motions in the desired direction or repeating the same pointing gesture. Robot player P4 responded to P1 by stepping back and moving forward towards the traffic light button again as confirmation. Recognising this, P1 then assisted by pressing the button. In this manner, both parties intriguingly communicated through a blend of verbal and non-verbal exchanges.

\subsubsection{Displaying emotions}

Even though participants were not given any external communication modalities apart from the robot body -- consisting of a box and a pole -- six of the participants attempted to convey emotions through movements or sound. A common emotion exhibited by participants (n=5) was anxiety when pedestrians did not offer help during the role-play. This anxiety was manifested through frequent and intense shaking of the pole or twisting of the robot's body. The displayed emotion raised empathy among participants. P17, for instance, reflected on an instance where the robot player was impeded by an obstacle and energetically waved its pole towards her: \emph{`[…] it (the robot) seemed very anxious. Then I quickly realised that it was this thing that was blocking it.'}
As a result, three out of the five participants who initially didn't assist the robot began to pay heightened attention and acknowledged the situation. This ultimately prompted them to step in and provide assistance. 

Four robot players tried to express gratitude after receiving help by expressing joyful emotions through their movements, such as hopping up and down (P6) or spinning around (P15), as well as through sound, like vocalising an uplifting tune (P7, P1). Displaying these joyful emotions prompted responses such as nodding (P9) and waving (P17, as shown in Fig.~\ref{Reactions}, right) by pedestrian players. P17 drew a connection between waving to the robot and her experience of encountering small animals, explaining, \emph{`Because I tend to greet small animals, or things that I find cute or have emotions.'} Furthermore, P7 conveyed a sense of satisfaction when seeing the robot spinning around, stating, \emph{`I felt very satisfied because I believe it has feelings. […] I help it and it is happy, which also makes me happy.'} P9, commenting on the joyful tune produced by the robot player P11 following his assistance, noted, \emph{`It made me feel like, alright, I did the right thing.'} 
\begin{figure*}[h]
\begin{center}
\includegraphics[width=1\textwidth]{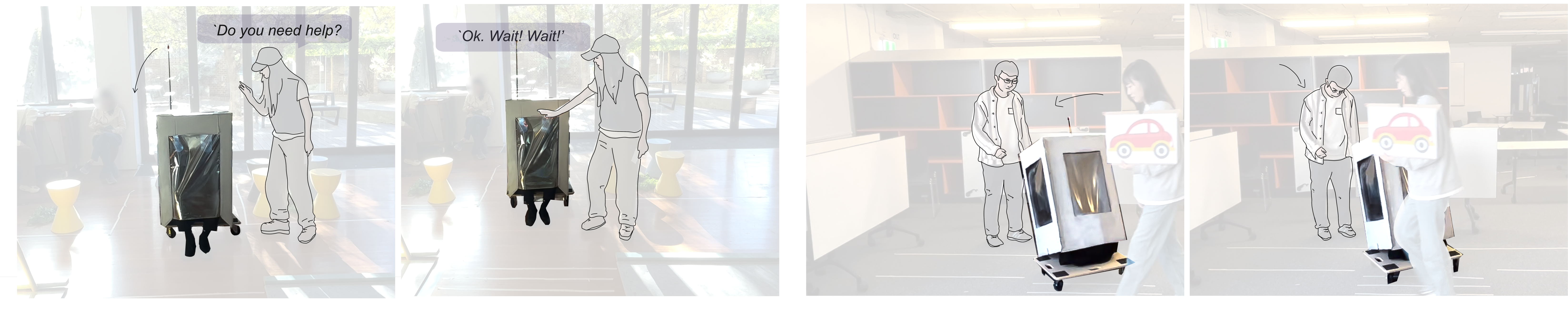}
\end{center}
\vspace{-5pt}
\caption{Communication between robot and pedestrian players: P8 inquires about robot player P6's needs, while P6 responds by pointing with the pole (left); Robot player P14 adopts a `bow' gesture towards the pedestrian player, eliciting a corresponding bow in return (right).}\label{bow}
\Description{Figure 4 showcases two screenshots depicting communication between robot and pedestrian players in a bodystorming session. In the first image on the left, Pedestrian player P8 is seen inquiring about Robot player P6's needs, with speech bubbles reading 'Do you need help?' and 'Ok. Wait! Wait!' The Robot player responds by pointing with a pole. In the second image on the right, Robot player P14 is shown adopting a 'bow' gesture towards the pedestrian player, which prompts the pedestrian to bow in return, demonstrating a reciprocal non-verbal communication exchange.}
\end{figure*}
\subsubsection{Demonstrating repetitive patterns}
Having discussed the three primary components of communications — addressing bystanders, cueing intentions, and displaying emotions — another notable observation emerged. Robot players frequently assembled these components into discernible repetitive patterns, resembling the predictable and programmed behaviours generally associated with robots (n=7).

P7, for instance, developed a unique routine to signify a pathway obstructed by an obstacle: she advanced towards obstacles while emitting two flat-tone beeps, moved back with an up-tone beep, and paused briefly before repeating this cycle multiple times. Her rhythmic auditory cues were synchronised with her physical movements. She later explained that this use of repetitive movements and audio cues was reflective of her \emph{`imagination of the robot having some program behind the system.'} Similarly, P2 adopted a pattern that combined cueing intention and addressing bystanders by repeatedly turning towards the pedestrian player, returning to the original position, and then turning towards the obstacles blocking its path.

Eight participants indicated that the recognition of programmed machine-like behaviour augmented their understanding of a robot's intent to communicate. In contrast, movements lacking a recognisable pattern were sometimes perceived as \emph{`erratic'} or \emph{`malfunctioning'}. The repetitive movement patterns reminded four participants of situations where domestic cleaning robots get stuck and repeatedly attempt to move back and forth. The familiar motion patterns helped participants form associations and understand the robot's need for help. 

In addition to enhancing understanding, recognising repetitive patterns in the robot's behaviour also potentially improved participants' confidence in the robot's abilities. The robot's consistent adherence to certain rules communicated a sense of control over its actions, as P9 noted, \emph{`I think the robot knew what it wanted.'} P7 indicated that the repetitive patterns in the robot's movements signify predictability, allowing \emph{`the pedestrian (to) anticipate what’s gonna happen.'}

\subsection{Factors shaping bystander decision to offer or decline assistance}

\subsubsection{Preconceptions of agent autonomy}

Our study underscores the prevailing perception among participants that service robots should operate with complete self-sufficiency and efficiency~(n=10). This forms a major reason for participants' reluctance to assist robots. For instance, P9 shared his presumption about robots' capabilities to manage all tasks autonomously, stating, \emph{`I thought the robot was able to do everything itself.'} Such misconceptions can foster misunderstandings about the actual capabilities and needs of these robots, an aspect further highlighted when P9 continued, \emph{`[…] so I didn't realise the robot was asking me to help.'} This expectation subsequently instigated skepticism among six participants regarding the functional utility of service robots that require human intervention. This sentiment was articulated through comments like,\emph{`If you have to work for them, then what’s the point to have a robot'} (P6). P7 further noted a decline in trust due to the robot's need for help, contrasting it with her expectation of a service robot's role as a functioning entity, stating \emph{`As a working robot (i.e., service robot), they kind of made me feel like untrust(worthy). […] so they (have to) work perfectly.'} 

Interestingly, when debriefed about the actual scenario, there was a notable shift in the attitudes of four participants. They came to understand that the robots' challenges arose from external factors outside their capabilities (e.g., obstacles purposely placed by humans) rather than any inherent malfunctions. P4 underscored this realisation, remarking, \emph{`then it's the human's fault (for placing the obstacle)'}. The reassignment of responsibility for the robot's immobilisation not only improved participants' inclination to assist but also enhanced their empathy towards the robot's predicament. P9 summarised this change of mind, noting that in this case the robot is \emph{`in need of help rather than being needy'}. 

\subsubsection{Absence of responsibility}
The notion of being a mere bystander or pedestrian, devoid of any responsibility towards the enacted robots, emerged as a primary factor influencing the decision not to assist among ten participants. This sense of detachment made them reluctant to invest their time and effort in helping `something' they didn't feel accountable for. P14 particularly highlighted resistance to being perceived as \emph{`free labour'} for commercial enterprises, posing the question: \emph{`why should I spend my time helping something that is making a profit?'} However, they later nuanced this statement by adding: \emph{`If it (the robot) is for a non-profit purpose, then I might be inclined to help, even if it means me being a bit delayed.'} 

The concerns of getting entangled in potential troubles further discouraged five participants from offering help. P1 expressed this concern, stating \emph{`I am afraid of touching it and breaking things. […] It could cause trouble if we touch it.'}
 
\subsubsection{Unfamiliarity with robotic technology}

Seven participants expressed hesitation to assist the robot due to a perceived lack of expertise. They felt ill-equipped as \emph{`random pedestrians'} (P7) to provide assistance to the robot, a task they believed was best left to professionals, as indicated by two participants.

The unfamiliarity with robotic technology sparked safety concerns among six participants, hindering them from offering help. This was further corroborated by our observations of five participants who actively distanced themselves or avoided the robot when it approached them for assistance. This evasion stemmed from the uncertainty about potential risks linked to the robot's predicament, as P7 stated, \emph{`I don't know if it's a tiny little issue or if it's going to explode or something.'} 

\subsubsection{Intrinsic motivation: empathy and emotional responses}

Our interviews indicate that intrinsic motivation plays a compelling role in promoting bystanders to assist the robot, with feeling empathy being the primary motivator (n=8). Participants described the robot using terms like \emph{`depressing'}, \emph{`frustrated'}, and \emph{`helpless',} signifying their ability to infer the robot's emotional states through observation of its movements within given contexts. For example, P9 noted that observing the robot's body swaying in an appeal for help prompted an association with vulnerable individuals, stating \emph{`[…], so (it's) like a child needs help or an old person needs help'}. 

In addition to empathy, six participants reported experiencing a sense of emotional reward, capturing feelings of \emph{`fulfilment'}, \emph{`satisfaction'}, and \emph{`delight'}, following their actions to assist the robots. P2, for example, articulated this sentiment as, \emph{`You helped it and witnessed it moving forward, which brings you a sense of satisfaction.'} Moreover, the gratitude exhibited by the robot reportedly amplified these emotional rewards (n=4). 

However, it was also evident that some participants demonstrated a reduced level of empathy towards robots. Specifically, P1 drew comparisons with other entities, affirming readiness to \emph{`stop for a dog or a cat, but not for a robot.'} He justified this attitude with his belief that \emph{`you can't expect to treat a robot as a human or as a living animal.'}  In one session, P15, who assumed the role of a vehicle driver, further underscored this perspective by simulating a horn by knocking on the board prop and voicing their impatience by yelling at the robot. Their behaviour was justified by their assertion that the robot \emph{`cannot be viewed as human.'} 

\subsubsection{Extrinsic motivation: entertaining value and material reward}

Apart from intrinsic motivation, our interviews highlighted the role of extrinsic motivation, stemming from both tangible and intangible incentives, in fostering helping behaviours towards robots.

Six participants anticipated entertainment value from helping robots, viewing this form of intangible incentive as an additional incentive to offer assistance.
As articulated by P9, the incorporation of a \emph{`surprise element'} into the interaction could further \emph{`spark joy.'} P12 also mentioned the potential of \emph{`transforming it (assisting robots) into a more game-like experience'}. 

Furthermore, two participants felt that their prosocial actions should also yield tangible advantages for them. They suggested the introduction of material rewards, such as vouchers or discounts from the company that implements the robot, could further stimulate their willingness to assist. This perspective was rooted in their belief that as bystanders, their prosocial behaviour towards the robot was not directly \emph{`benefiting'} them.

\subsection{Desired robot characteristics}
Based on the results of the word sorting activity and insights derived from the participants' discussions, we identified three characteristics that help-seeking robots could implement, which we present in this section.

\subsubsection{Vibrantly mechanical}

The physical sensations experienced by robot players are evenly distributed among all words with a slightly higher number of \emph{`clumsy'} (n=4). This choice primarily stems from the physical limitations of moving within the robot costume.

There's a noticeable overlap between the perceived physical qualities of the robot player and the desired attributes. In particular, \emph{`mechanical'} was selected as perceived quality (n=9) and desired quality (n=10). Similarly, the same pattern was observed in \emph{`playful'} (8 for perceived quality, 14 for desired quality), and \emph{`cute'} (11 for perceived quality, 8 for desired quality). This correlation implies that the robot players' behavioural strategies, to some extent, met participants' expectations for how a robot should act when seeking assistance, especially concerning these attributes. 
The movement of the robot players, encapsulated in the minimalist abstract box-shaped costume, inherently conveys an impression of cuteness without the need for additional embellishments. As P12 expressed, \emph{`I feel its existence is cute enough already. Just imagine you're on the road, helping a little robot, and it goes "dibbly-dobbly" as it moves forward. It's already incredibly cute, and there's no need to add additional design language.'} While some participants also selected \emph{`lively'} and \emph{`organic'}, they view these as elements that can be added to the overall mechanical nature of the robot to enhance the expressiveness. P13 emphasised the importance of using these elements thoughtfully to \emph{`avoid creating the uncanny valley effect.'}

\subsubsection{Cheerfully confident}

In regards to the psychological feelings of robot players, there was no clear tendency being identified in the word sorting, which may imply that this was highly subjective. When it comes to other participants' perceptions of the robot player, \emph{`brave'} (n=9) and \emph{`confident'} (n=8) emerge as two dominant qualities. Participants highly commend the robot player's efforts to find solutions in challenging situations. P16 pointed out the robot's vulnerability and the hazards and difficulties of its environment. She articulated, \emph{`(the robot) dared to cross the road on his own and was thinking about how to do it.'} when it was \emph{`dangerous because there were no traffic lights for (it)'}. P2 contemplated the perceived braveness, projecting the psychological state that humans have when seeking help from strangers onto the robot. She expressed that the robot \emph{`needs help (and asks for it), as it needs to use courage to convey the help it needs.'} 

In terms of the desired psychological attributes, participants generally favoured positive emotions. The term \emph{`confidence'}~(n=12) emerged as the predominant expectation that people had for the robot's demeanour. P7 expressed this perspective by describing a service robot as \emph{`some kind of professional stuff'}, implying the need for the robot to exhibit characteristics that align with expected proficiency. In addition, participants generally expected the robot to display \emph{`happy'}~(n=6) emotions after receiving help.

Descriptors of negative emotional valence (e.g., \emph{`sad'}) were not considered as desired qualities for help-seeking robots. P2 offered an enlightening comment that a casually-encountered robot exhibiting sadness while seeking assistance was reminiscent of street beggars, leading to feelings of what they described as \emph{`emotional blackmail.'}

\subsubsection{Responsively outgoing}
The self-assessed feeling of \emph{`needy'} emerged as the most frequently chosen term among robot players~(n=7), indicating that they experienced a sense of helplessness and a perceived necessity for human aid as a robot in the given situation. To elicit help, most of the robot players opted to project a more approachable character, frequently choosing descriptors like \emph{`extrovert'}~(n=4), \emph{`collaborative'}~(n=4) and \emph{`inviting'}~(n=3). In contrast, only two robot participants resonated with terms like \emph{`shy'} and \emph{`introvert'}.

Other participants' perceived robot social quality aligned with the self-assessed social attribute of robot players, with \emph{`needy'} (n=15) and \emph{`extroverted'} (n=8) being the most frequently chosen terms.
In terms of the desired social qualities, although \emph{`extroverted'} was still among the preferred terms, participants placed a greater emphasis on communication qualities such as being \emph{`polite'}, \emph{`responsive'} and \emph{`collaborative'}. P2 highlighted the importance of politeness, even when the robot is in urgent need of help, saying, \emph{`At the same time, when you try to attract everyone's attention as much as possible, you also need to be gentle towards others.'} The desired \emph{`responsive'} quality reflects participants' expectation of receiving feedback after offering help. For example, P17 expressed feeling disappointed if she helped the robot without receiving any response from it. 

\vspace{1pt}
\section{Discussion}
Drawing from the findings of the previous section, we identify three design considerations (C1-3) for fostering casual human-agent collaboration. In addition, we reflect on the strengths and limitations of the bodystorming design activity employed in our study. 

\subsection{Expressiveness through functionality-oriented form}

Given humans' innate psychological tendency to interpret social cues from moving objects~\cite{heider1944experimental}, physical movement has become a pivotal medium in human-agent interaction to facilitate social communication ~\cite{Zuckerman2015conversational,Zaga2017nod, Petra2017MovementMatters, Luria2017Comparing,Anderson-Bashan2018Greeting}. An example of extreme abstraction with minimal movement serving as social cues is the \emph{`Greeting Machine'}~\cite {Anderson-Bashan2018Greeting} – a small ball rolling on a bigger dome in varied trajectories. This design effectively elicits both positive and negative social encounter responses. Similar to \emph{`Greeting Machine'}, participants in our design investigation were constrained by a robotic costume made of a box and a pole that had only minimum expressive capabilities. This costume had no extensions beyond the basic form factor of a conventional delivery robot that is primarily intended for the task of transporting goods. Nonetheless, subtle cues, such as the robot's orientation in its shape (i.e. orienting the front of the box costume toward an object) or the directionality of simple components (i.e., pointing in the intended direction using the pole), proved effective in addressing bystanders and conveying the robot's need for assistance. This was evident, as in all the sessions, pedestrian players accurately understood the robot player's request for assistance. In addition, we even witnessed conversations formulated through the back-and-forth interplay between pedestrian inquiries and the subtle motions of robot players (as shown in Fig.~\ref{bow}, left)

In addition to empathising with the robot's feelings of frustration and vulnerability when seeking help, it was evident that people also recognise social signals gestured through subtle movements, such as gratitude. For instance, a simple tilt of the box-shaped robot body, can be perceived as a `bow', even prompting the pedestrian player P15 to bow back (as shown in Fig.~\ref{bow}, right).



As pointed out in our literature review, linguistic utterances have been central in human-agent collaboration~\cite{Seaborn2021voice} and represent a key method for help-seeking requests~\cite{Backhaus2018HelpFrameWork,Srinivasan2016Politeness,Huttenrauch2003ToHelpOrNot} given their effectiveness in conveying information. Nonetheless, challenges such as cultural and language barriers, cognitive load, and issues related to noise and distortion in public settings limit their utility in the context of casual human-agent collaboration in public spaces. In addition to that, our study revealed further concerns regarding the appropriateness of agents verbally asking for help in urban public contexts (n=7). P1, for instance, suggested it could be a \emph{`bit abrupt or out of place'} if a robot suddenly started to talk human language on the street, underscoring the need for robots to have their own unique, natural communication modes. Additionally, safety concerns were expressed by three participants who suggested that language used by robots might potentially distract other road users. Notably, P14 expressed potential discomfort in feeling exploited by commercial entities when robots use human language to solicit assistance, suggesting a \emph{`feeling of being used as free labour for those commercial companies'}. 


\vspace{-6pt}
\begin{itemize}[label={},leftmargin=0pt]
\setlength\itemsep{0.5em}

\begin{snugshade*}
\item 
\begin{minipage}{\linewidth-2\fboxsep-2\fboxrule} 
  \vspace{4pt} 
\textbf{C1} - The design of agent help-seeking strategies should leverage the inherent expressiveness found in the functional aspects or form of the agent. While ensuring effective initiation of help-seeking requests, these implicit communication channels can prevent from being viewed as disruptive.
 \vspace{4pt} 
\end{minipage}
\end{snugshade*}

\end{itemize}

\subsection{Adherence to perceived agent social categories}

Robots and intelligent agents, growing rapidly in sophistication and sociability, have spurred enhanced research into agent social identity~\cite{hogg2016SIT}, delving into aspects like gender~\cite{eyssel2012Gender}, age~\cite{EDWARDS2019Age}, and race~\cite{Bartneck2018Race}. This trajectory was echoed at a recent workshop~\cite{winkle2021Identity}, which emphasised the importance of designing robots that can effectively convey social identities to optimise human-robot interaction outcomes. While this workshop primarily centred on social identities associated with specific attributes (i.e. gender), our findings broaden this scope to encompass wider social categories~(e.g. occupations), which should be considered in the design of casual collaborations between agents and bystanders. 


In our investigation of the help-seeking delivery robot, participants perceived these robots primarily as service providers or professional workers. This perception led them to expect high proficiency from these robots, favoring the robot presenting qualities reminiscent of confidence over neediness. Consequently, even when assistance was necessary, participants displayed a general reluctance to interact with robots that seemed overly needy or showcased negative emotions. This inclination stands in contrast with studies on eliciting human prosocial behaviour towards social companion robots~\cite{Connolly2020Prosocial, Daly2020RobotInNeed} (e.g., robotic pets). In these settings, negative emotional expressions in robots often serve as a catalyst, motivating individuals to step in and offer help. This difference could be rooted in the distinct perceived categories: \emph{`worker'} versus \emph{`companion'}. 
In public urban settings, intelligent agents participate integrally in various facets of urban life. As a result, they embody a wide range of social categories, from service providers (e.g., delivery robots~\cite{GEHRKE2023SideWalk}, street cleaning robots~\cite{Jeongmin2017Sweep}) and authoritative entities (e.g., smart traffic regulators~\cite{Lee2022}, patrol robots~\cite{Konrad2022Police}) to street entertainers (e.g., playful urban robots~\cite{Marius2020Woody, Lee2020Bubble}).
\begin{itemize}[label={},leftmargin=0pt]
\setlength\itemsep{0.5em}

\begin{snugshade*}
\item 
\begin{minipage}{\linewidth-2\fboxsep-2\fboxrule} 
  \vspace{5pt} 

\textbf{C2} - To respond to real-world expectations and social norms, the design of agent help-seeking strategies should adhere to and match with their perceived social categories~(e.g., occupation).

 \vspace{5pt} 
\end{minipage}
\end{snugshade*}

\end{itemize}


\subsection{Curating incentives: material rewards, act of care, or playful engagement} 

Though in 7 out of 9 sessions, pedestrian participants offered help, 4 of them mentioned they might not behave the same way in real-life settings. Furthermore, the misaligned objectives and the imbalanced benefits between agents and bystanders in such casual collaborations highlight the need for incentives.

Previous research on bystander assistance for commercially-deployed robots has divided `help' into two broader categories: either `helping-as-work', emphasising precarity and invisible labour, or `helping-as-care', which accentuates the emotional and relational dynamics of help~\cite{hakli2023helpingAsWorkCare}.
This work sheds light on the inherent ambiguity of these helping behaviours and prompts a rethinking of robot design to better shape these engagements. 
Our focus group findings resonate with this notion, while also shedding light on how different perspectives on helping robots call for various forms of incentives.





Offering material rewards, such as vouchers or discounts from companies deploying robots, turns casual collaborations into mutually beneficial exchanges. This model of paid crowdsourcing, evident in cases like identifying shared bicycle locations~\cite{griffin2019crowdsourcing} or urban data collection on platforms like OpenStreetMap~\cite{crooks2015crowdsourcing}, could be adapted for interactions with intelligent agents in public spaces. By framing casual collaborations as beneficial exchanges through material rewards, it can effectively align disparate objectives between parties into actions that are mutually advantageous.

In our study, empathy -- as an act of care -- emerged as a predominant motivator for offering assistance, manifested as a form of internal incentive. Empathy, essential in shaping communication and social bonds, has been underscored as a central component in human-agent interactions~\cite{Bickmore2005Empathy}. A robust body of research validates robots' capacity to elicit empathetic responses from humans~\cite{Kwak2013empathize,Riek2009Empathizing, rudovic2018personalized}. Correspondingly, studies on social robots have shown that these forms of empathy can drive prosocial behaviour, compelling humans to intervene against robot mistreatment~\cite{Connolly2020bully} or engage in affectionate actions, like petting~\cite{Heerink2012child}. Our findings resonate with these prior studies, suggesting that individuals exhibit empathy also towards public agents in need, driven by the observation of context and agent expressions. 

In addition to viewing helping robots as a form of work or act of care, `helping-as-play' has emerged as another perspective in our findings, for which the resulting entertaining value can function as an incentive.
Playful strategies have been used in various human-agent interaction settings, such as motivating children's learning~\cite{ahtinen2020learning} or encouraging factory workers in collaboration with robots~\cite{Chowdhury2021PlayHRC}. Furthermore, it has been shown to effectively engage the public and generate enjoyable experiences among bystanders ~\cite{Lee2020Bubble, Marius2020Woody}. 
One of the primary challenges for urban robots to ask for help from bystanders is convincing them to invest their time and tolerate potential disruptions. The inherent playfulness in humans could potentially act as an incentive for casual collaborations by transforming disruptions into pleasure and enjoyment. That being said, `helping-as-play' also needs to be carefully employed, as it can present ethical concerns similar to those in `helping-as-care'~\cite{hakli2023helping}.

\begin{itemize}[label={},leftmargin=0pt]
\setlength\itemsep{0.5em}

\begin{snugshade*}
\item 
\begin{minipage}{\linewidth-2\fboxsep-2\fboxrule} 
  \vspace{5pt} 
%
\textbf{C3} - To compensate for the misaligned task objectives in casual human-agent collaboration, the design of agent help-seeking strategies should incorporate appropriate incentives, transforming assistive behaviours into experiences that benefit both parties.

 \vspace{5pt} 
\end{minipage}
\end{snugshade*}

\end{itemize}

\subsection{Reflections and limitations of bodystorming design activity}

The physical constraints introduced by the robot costume, such as limited field of view, changes in perspective, and restricted mobility, while not capable of entirely replicating a robot's perspective, facilitated participants in departing from a conventional human viewpoint and immersing themselves in the sensations of robotic alienation and otherness ~\cite{Judith2020BecomingARobot}. As articulated by P5, \emph{`There's a sense of feeling out of place or not quite fitting in. It seems like there are no peers of my kind in the surroundings. Everyone else is tall, and I am short, so I feel a bit out of place or different.'} This sense of otherness could contribute to participants' emotional engagement when they assumed the robot role, which was evident in the frequently conveying sentiments of \emph{`frustration'} or \emph{`depression'}.  Additionally, beyond mere empathy with the robot's emotional state, the robot players also displayed a recognition of its societal function (i.e. its duty as a delivery service entity). For instance, P4 suggested feelings of \emph{`motivated'} and \emph{`happy'} when he \emph{`could continue working'}. This profound resonance with the agent's perspective underscores that our bodystorming approach effectively incorporated the agent's perspective into the design exploration. However, it is worth noting that, despite the effectiveness of the physical probes in consciously shifting participant's sensations towards the perspective of an agent~\cite{Judith2020BecomingARobot}, it is impossible to completely transcend the human standpoint through these methods as our human nature inherently separates us from things~\cite{Spiel2020}.


Consequently, the `participation' of agents in the design process surfaces tensions between humans and agents that stem from the misalignment in goals and benefits in a casual collaboration encounter. P9's behaviours while playing the robot role offer a striking example. He assertively used the robot costume to brush up against the pedestrian player's chest — a distinctly pushy gesture meant to force attention and assistance. He later admitted, ~\emph{`I was about to give up being nice'}, and even pondered, ~\emph{`That's why I was considering adopting a more aggressive strategy. I thought, I might just try pushing him onto the road.'}. This elicitation of tension and physical friction is less likely to surface in methods that take a purely human-centred perspective or those that rely solely on cognitive abilities. Furthermore, this tension was leveraged to stimulate design ideation, exemplified by P9's subsequent ideas that emerged from this physical friction. He suggested infusing robots with seemingly annoying nudging behaviours and \emph{`creating entertainment value'} to possibly uplift people's moods and thus promote pro-social behaviour. 

Despite the strengths of our approach in generating design insights, we acknowledge its limitations. Discussions and reactions concerning helping a casually encountered agent were elicited by role-playing activities, without the incorporation of real technology. Even though the immersive nature of the replicated scenario and participant engagement is evident, the absence of real intelligent agents and the artificial nature of scenarios replicated in laboratory settings might limit the direct applicability of our findings. Acknowledging these inherent limitations of bodystorming method, future research should validate and refine our design considerations through more spontaneous interactions between humans and high-fidelity artifacts (e.g., through Wizard of Oz testing~\cite{Wizard1993} or virtual reality study~\cite{VR2018}).




\section{Conclusion}

Intelligent agents are increasingly transitioning into uncontrolled settings that often may challenge their operational capabilities. While collaboration between humans and agents can offer means to overcome these challenges, to date little is known about how to create effective and engaging human-agent collaboration in casual settings~(e.g., involving surrounding bystanders). 

Through employing bodystorming and real-world help-seeking scenarios encountered by urban robots, we uncovered potential help-seeking strategies through participants' enactments as robots, including addressing bystanders to initiate interaction and using non-verbal cues to communicate intention and request. Furthermore, the display of emotions and demonstrating repetitive patterns has been found to ease help-seeking requests.
Taking into consideration the reactions of pedestrian players from our bodystorming activities, we further offered insights into the factors influencing people's response to robot's help-seeking requests and identified desired robot characteristics. 

Synthesising insights from both robot and bystander perspectives, we conclude with a set of design considerations for help-seeking agents that operate in uncontrolled environments. These considerations include promoting expressiveness, ensuring alignment with agent social categories, and curating appropriate incentives. Findings from our design exploration and considerations for implementing help-seeking strategies provide a foundation for creating intelligent agents that operate in uncontrolled environments and aim to facilitate casual collaboration that is mutually beneficial to humans and agents.
\begin{acks}
This study is funded by the Australian Research Council through the ARC Discovery Project DP220102019 Shared-Space Interactions Between People and Autonomous Vehicles. We thank all the participants for taking part in this research. We also thank the anonymous CHI'24 reviewers and ACs for their constructive feedback and suggestions to make this contribution stronger.   
\end{acks}

\balance{}
\balance{}

\bibliographystyle{ACM-Reference-Format}
\bibliography{Help}

\end{document}